\documentclass[11pt,a4paper]{iopart}

\usepackage{graphicx}

\begin{document}

\title[Magnetism in Mg$_{1-x}$Cu$_x$Cr$_2$O$_4$]
{Evolution of magnetic properties in the normal spinel solid solution 
Mg$_{1-x}$Cu$_x$Cr$_2$O$_4$}

\author{Moureen C. Kemei, Stephanie L. Moffitt, and Ram Seshadri}
\address{Materials Department, University of California, 
Santa Barbara CA 93106}
\eads{kemei@mrl.ucsb.edu}

\author{Daniel P. Shoemaker}
\address{Material Science Division, Argonne National Laboratory, 
Argonne IL 60439}

\begin{abstract}

We examine the evolution of magnetic properties in the normal spinel oxides 
Mg$_{1-x}$Cu$_x$Cr$_2$O$_4$ using magnetization and heat capacity measurements.
The end-member compounds of the solid solution series have been studied in 
some detail because of their very interesting magnetic behavior. 
MgCr$_2$O$_4$ is a highly frustrated system that undergoes a first order 
structural transition at its antiferromagnetic ordering temperature. 
CuCr$_2$O$_4$ is tetragonal at room temperature as a result of 
Jahn-Teller active tetrahedral Cu$^{2+}$ and undergoes a magnetic transition 
at 135\,K. Substitution of magnetic cations for diamagnetic Mg$^{2+}$
on the tetrahedral \textit{A} site in the compositional 
series Mg$_{1-x}$Cu$_x$Cr$_2$O$_4$ dramatically affects magnetic behavior. In 
the composition range 0 $\leq$ $x$ $\leq$ $\approx$0.3, the compounds are 
antiferromagnetic. A sharp peak observed at 12.5\,K in the heat capacity of
MgCr$_2$O$_4$ corresponding to a magnetically driven first order structural
transition is suppressed even for small $x$ suggesting glassy disorder. 
Uncompensated magnetism -- with open magnetization loops 
-- develops for samples in the $x$ range $\approx0.43$ $\leq$ $x$ $\leq$ 1.
Multiple magnetic ordering temperatures and large coercive fields emerge in
the intermediate composition range $0.43 \leq x \leq 0.47$. 
The N\'{e}el temperature increases with increasing $x$ across the series 
while the value of the Curie-Weiss $\Theta_{CW}$ decreases. A magnetic
temperature-composition phase diagram of the solid solution series is
presented.

\pacs{
     75.30.Kz %Magnetic phase boundaries (including magnetic transitions, 
              % metamagnetism, etc.) 
	 75.47.Lx %Magnetic oxides
     75.50.Ee %Antiferromagnets
     75.50.Gg %Ferrimagnets
     }
\end{abstract}

\maketitle

\section{Introduction}

Materials with the spinel structure display a wide range of functional
properties and are applied as battery electrodes \cite{Thackeray1987},
multiferroic materials\cite{Yamasaki2006,Lawes2006}, and catalytic
materials.\cite{Gao2009} In addition, spinels offer unique opportunities for
the exploration of exotic magnetic
phenomena.\cite{Yaresko2008,Loidl2009,Takagi2005} A rich diversity of
possible magnetic ground states can be found in materials with the spinel
structure. These range from long range ordered ferrimagnetic states,
observed in magnetite Fe$_3$O$_4$, to degenerate spin
liquid states seen in systems with geometrically frustrated interactions
such as ZnCr$_2$O$_4$.\cite{Cheong2002} In spinels with the general formula
\textit{AB}$_2$O$_4$, cation \textit{A} and \textit{B} sites can both be 
occupied by magnetic ions. Coupling between the various magnetic ions gives 
rise to a number of competing exchange pathways and a multitude of possible 
ground states. Additional complexity arises when antiferromagnetically coupled
cations populate the pyrochlore \textit{B} sublattice in spinels. In this
configuration, geometric constraints preclude the realization of a unique
ground state resulting in frustrated systems.  Slight perturbations of
highly degenerate spin liquid states in frustrated magnets can result in a
range of novel behavior.\cite{Balents2010, Ramirez2006}   
 
Strongly frustrated three-dimensional pyrochlore \textit{B} sublattices
occur in oxide spinels with a non-magnetic \textit{A} site and a chromium
\textit{B} site. Cr$^{3+}$ with a [Ar]3d$^3$ electron configuration
shows a preference for the octahedral site where crystal field splitting
stabilizes a half filled $t_{2g}$ level separated in energy from an
empty $e_g$ level.\cite{Orgel1957,Miller1959}
Antiferromagnetic nearest neighbour (NN) interactions between Cr$^{3+}$ ions
populating the pyrochlore \textit{B} sublattice cannot be fully satisfied.
Pyrochlore sublattices with antiferromagnetically coupled spins have been
shown to result in frustrated Heisenberg spin Hamiltonian
expressions where NN interactions alone would not result in a single low
energy state even at $T$ = 0\,K.\cite{Anderson1956,Cheong2007,Chalker1997,
Sondhi2002} Given the strongly frustrated Cr$^{3+}$ sublattice, the choice
of \textit{A} site cation can profoundly affect magnetic properties in
chromium oxide spinels.  

Magnetic ground states depend strongly on the \textit{A} cation in
\textit{A}Cr$_2$O$_4$ spinels.  Non-magnetic \textit{A} site (for instance
\textit{A} = Zn, Mg, Cd) chromium oxide spinels are highly
frustrated.\cite{Loidl2010} Spin-lattice coupling resolves the large ground
state degeneracy by selecting a unique ordered state \textit{via} a spin 
Jahn-Teller effect at the magnetic ordering temperature.\cite{Sondhi2002} In 
magnetic \textit{A} site spinels (for instance \textit{A} = Co, Fe, Cu, Mn),
\textit{A}-O-Cr coupling dominates over frustrated Cr-Cr interactions and
non-collinear ferrimagnetic ground states are 
attained.\cite{Seshadri2006, Ederer2007, Prince1957, Pickart1964, Suzuki2004, Tokura2009} Understanding changes in 
interactions due to gradual addition of magnetic ions on the \textit{A} site 
of frustrated \textit{A}Cr$_2$O$_4$ spinels is important. In this study, we 
investigate the magnetic properties of the solid solution 
Mg$_{1-x}$Cu$_x$Cr$_2$O$_4$ where the end members MgCr$_2$O$_4$ and 
CuCr$_2$O$_4$ differ both structurally and magnetically.

The canonical geometrically frustrated spinel MgCr$_2$O$_4$ crystallizes in
the cubic space group $Fd\bar3m$ at 300\,K. The
pyrochlore Cr$^{3+}$ sublattice is based on a triangular motif where
antiferromagnetic NN Cr-Cr coupling is geometrically frustrated. As a
result, the spins in MgCr$_2$O$_4$ remain disordered far below the
theoretical ordering temperature 
($\Theta_{CW} \approx -400$\,K).\cite{Tsunoda2007,Cava2011} 
A structural distortion lifts the spin state
degeneracy of the pyrochlore Cr$^{3+}$ sublattice at the N\'{e}el
temperature ($T_N$) $\approx$ 12.5\,K.  The tetragonal space group
$I4_1/amd$ has been suggested as the low symmetry crystallographic
structure.\cite{Attfield2008,Baehtz2002} A sharp peak in heat capacity
coincident with the magnetically driven structural transition in MgCr$_2$O$_4$ 
denotes the first-order nature of this transition.\cite{Gmelin2000} 

The spinel CuCr$_2$O$_4$ is cubic above 873\,K,\cite{Kanamori1960,Murthy1983} 
with Cu$^{2+}$ occupying tetrahedral sites and Cr$^{3+}$
populating octahedral sites. In the cubic phase, tetrahedral crystal field
splitting of the d$^9$ Cu$^{2+}$ energy level results in triply
degenerate high lying $t_2$  subshells and fully occupied low
energy $e_g$ subshells.\cite{Gerloch1981} Distortion of the CuO$_4$ tetrahedra 
lifts the orbital degeneracy of the $t_2$ level, and when these distortions 
are coherent, the symmetry of CuCr$_2$O$_4$ is lowered from cubic $Fd\bar3m$ to 
tetragonal $I4_1/amd$.\cite{O'Neill1997} Magnetic studies of CuCr$_2$O$_4$
show that it is ferrimagnetic at 135\,K.  The magnetic structure in the
ordered state is triangular with Cr$^{3+}$ in the 001 planes alligned
parallel but at an angle to Cr$^{3+}$ in adjacent planes. Cu$^{2+}$ align
antiparallel to the net moment of the Cr$^{3+}$ sublattices forming a
magnetic structure comprising triangles of spins.\cite{Prince1957}

Previous work has investigated the local and average structural changes in
the series Mg$_{1-x}$Cu$_x$Cr$_2$O$_4$.\cite{Shoemaker2010}  These studies
show that although local distortions occur for all Cu$^{2+}$ substituted
compositions, cooperative structural changes are dependent on $x$ 
and on temperature. For example, at room temperature, compounds of the series 
Mg$_{1-x}$Cu$_x$Cr$_2$O$_4$ remain cubic on average for $x$ $<$ 0.43. 
Tetragonal average symmetry driven by cooperative Cu$^{2+}$ Jahn-Teller 
distortions appears in compositions with $x$ $>$ 0.43 at 300\,K. 

We study the effect of introducing magnetic, Jahn-Teller active Cu$^{2+}$ on 
the non-magnetic \textit{A} site of MgCr$_2$O$_4$ on magnetic frustration and 
on the nature of magnetic interactions. A previous study of the system
Zn$_{1-x}$Co$_x$Cr$_2$O$_4$ showed that addition of Co$^{2+}$ on the
non-magnetic Zn$^{2+}$ site quenched frustration across the
series.\cite{Melot2009} Here, we explore changes in magnetic behavior as
Jahn-Teller active tetrahedral Cu$^{2+}$ induces lattice distortions in
addition to adding magnetism on \textit{A} site of the compounds
Mg$_{1-x}$Cu$_x$Cr$_2$O$_4$.\cite{Shoemaker2010} Novel properties such as
intrinsic exchange bias have been shown to exist at phase
boundaries.\cite{Shoemaker2009} We therefore closely examine the region
between the antiferromagnetic and ferrimagnetic phases in
Mg$_{1-x}$Cu$_x$Cr$_2$O$_4$ for unusual phenomena.   

\section{Experimental details}

Polycrystalline samples of the series Mg$_{1-x}$Cu$_x$Cr$_2$O$_4$
($x$ = 0, 0.1, 0.2, 0.43, 0.47, 0.6, 0.8, and 1) were prepared
through calcination of nitrate precursors as reported by Shoemaker
and Seshadri.\cite{Shoemaker2010} The samples were structurally
characterized by laboratory x-ray diffraction using a Philips X'pert
diffractometer with Cu-K$\alpha$ radiation. Phase purity was
confirmed in selected compositions using high-resolution synchrotron powder
x-ray diffraction collected at the 11-BM beamline at the Advanced Photon
Source. Some of the samples have also been previously characterized by
time-of-flight neutron scattering which verified that all the spinels
are normal, meaning the \textit{B} site is purely Cr$^{3+}$, and there is
not Cr$^{3+}$ on the \textit{A} site. Magnetic susceptibility measurements on
powder samples were performed using a Quantum Design MPMS superconducting
quantum interference device (SQUID) magnetometer. In all samples,
magnetization as a function of temperature was measured at a field of 0.1\,T.
Isothermal field dependent magnetization measurements were performed at 2\,K
and 5\,K. A Quantum Design Physical Properties Measurement System was used to
measure heat capacity at 0\,T for various temperature ranges selected to
accommodate the magnetic transition temperature of each sample. For the heat
capacity measurements, pellets of 50\,\% sample mass and 50\,\% silver mass
were prepared by grinding and pressing at about 330\,MPa. Silver was used to
increase mechanical strength and thermal conductivity. The pellets were 
mounted on the sample stage using Apiezon N grease. The heat capacity of
the grease and of silver heat capacity were collected separately and 
subtracted to obtain the sample heat capacity. 
	
\section{Results and Discussion} 

\begin{figure}
\centering 
\includegraphics[width=0.75\textwidth]{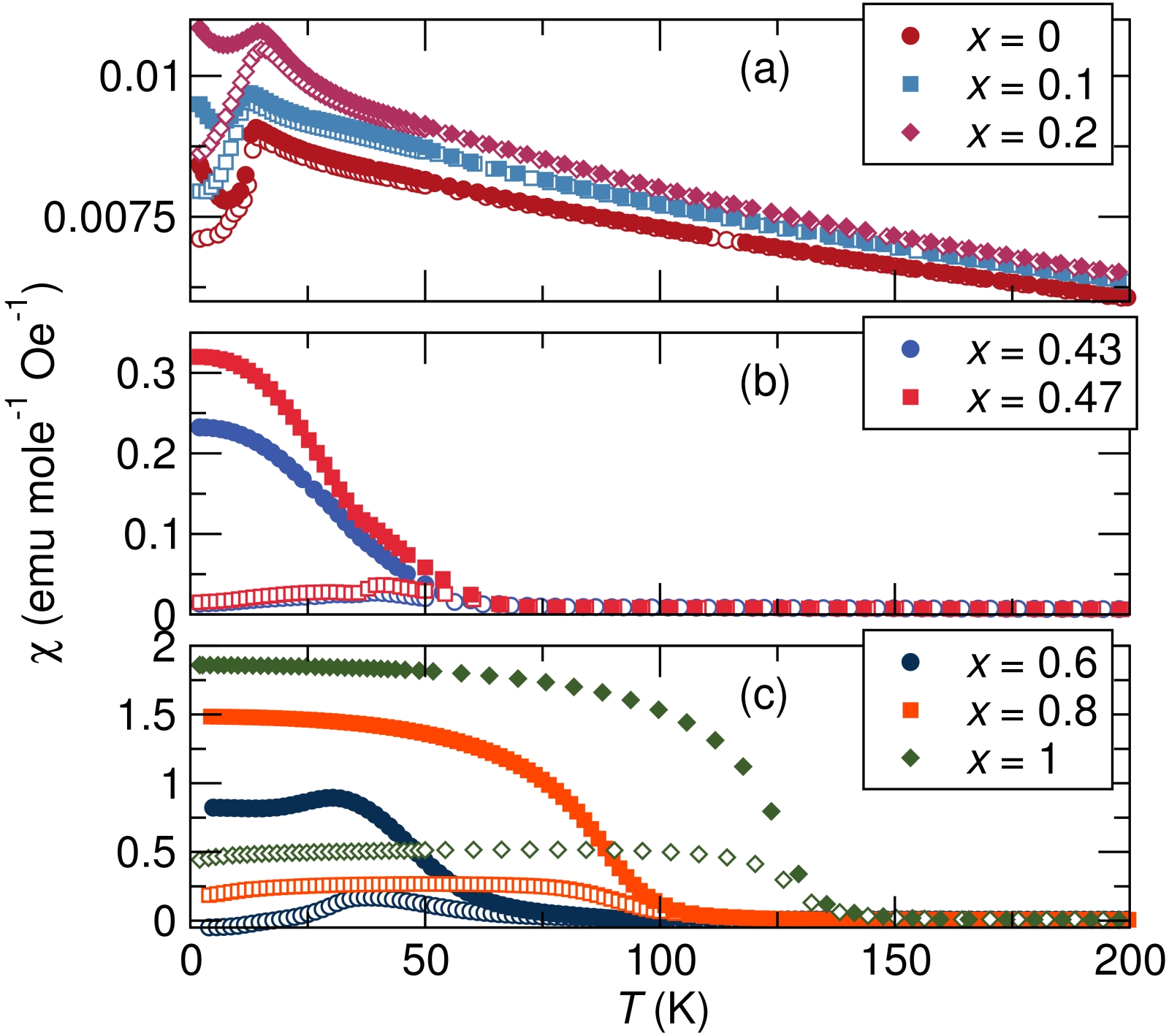} 
\caption{Zero-field-cooled (open symbols) and field-coled (closed symbols)
magnetic susceptibility as a function of temperature of the series
Mg$_{1-x}$Cu$_x$Cr$_2$O$_4$ under a DC fields of 0.1\,T (a) samples with 
$x$ $<$ 0.4, (b) Samples with 0.4 $<$ $x$ $<$ 0.5, and (c) Samples 
$x$ $>$ 0.5. Magnetic ordering temperatures and the magnetization at low 
temperatures increase with $x$ concentration.\label{fig:chi}}
\end{figure}

We study the magnetic properties of the compounds Mg$_{1-x}$Cu$_x$Cr$_2$O$_4$ 
using magnetic susceptibity measurements. Zero
field cooled (ZFC) and field cooled (FC) temperature dependent magnetic
susceptibilities of the system show a composition dependent ordering
temperature ($T_{N}$) defined as the temperature where d$\chi$/d$\it{T}$ is
maximum (Figure\,\ref{fig:chi} and Table\,\ref{table:magnetism}). $T_{N}$
increases with Cu$^{2+}$ substitution. In addition to Cr$^{3+}$-Cr$^{3+}$
interactions that are present in all compositions, Cu$^{2+}$-Cr$^{3+}$ and
Cu$^{2+}$-Cu$^{2+}$ interactions occur in copper doped samples. The increase
in $T_N$ correlates well with the increase in the number of magnetic
interactions. It is also likely that Cu$^{2+}$ moments compensate some of
the Cr$^{3+}$ sublattice moments. As Cr$^{3+}$ are compensated, the
difficulty involved with satisfying antiferrromagnetic interactions between
spins on a pyrochlore lattice is alleviated easing frustration and allowing
magnetic order at high $T_N$. The sample $x$ = 0.1 shows a lower
$T_N$ suggesting that randomly distributed dilute Cu$^{2+}$ ions may be
disrupting long range order, causing the system to freeze into a disordered
spin state at lower temperature. The decrease in $T_N$ with dilute doping
of magnetic cations on the \textit{A} site of a frustrated antiferromagnet
has been observed in the system Zn$_{1-x}$Co$_x$Cr$_2$O$_4$.\cite{Melot2009}

In samples $x$ = 0, 0.1, and 0.2 susceptibility increases to a
maximum at $T_N$ before decreasing below $T_N$ [Figure\,\ref{fig:chi}(c)]. 
This trend in both ZFC and FC data indicates  dominant 
antiferromagnetic interactions. A sharp cusp is observed at $T_N$ in
MgCr$_2$O$_4$ where long range magnetic order occurs \textit{via} a spin-driven
Jahn-Teller transition.\cite{Broholm2010} This cusp broadens in samples
$x$ = 0.1 and 0.2 indicating changes in the magnetic ground state
with Cu$^{2+}$ doping. Low Cu$^{2+}$ content may contribute to glassy
behavior by introducing asymmetric exchange pathways throughout the sample.
The decrease in ZFC susceptibility coupled with the increase in FC
susceptibility below $T_N$ in compositions $x$ = 0.43, 0.47, 0.6, 0.8
and 1 [Figure\,\ref{fig:chi}(a) and (b)] demonstrates ferrimagnetic behavior.
The magnitude of the low temperature susceptibility increases with Cu$^{2+}$
content as shown in Figure\,\ref{fig:chi}.

\begin{figure}
\centering 
\includegraphics[width=0.75\textwidth]{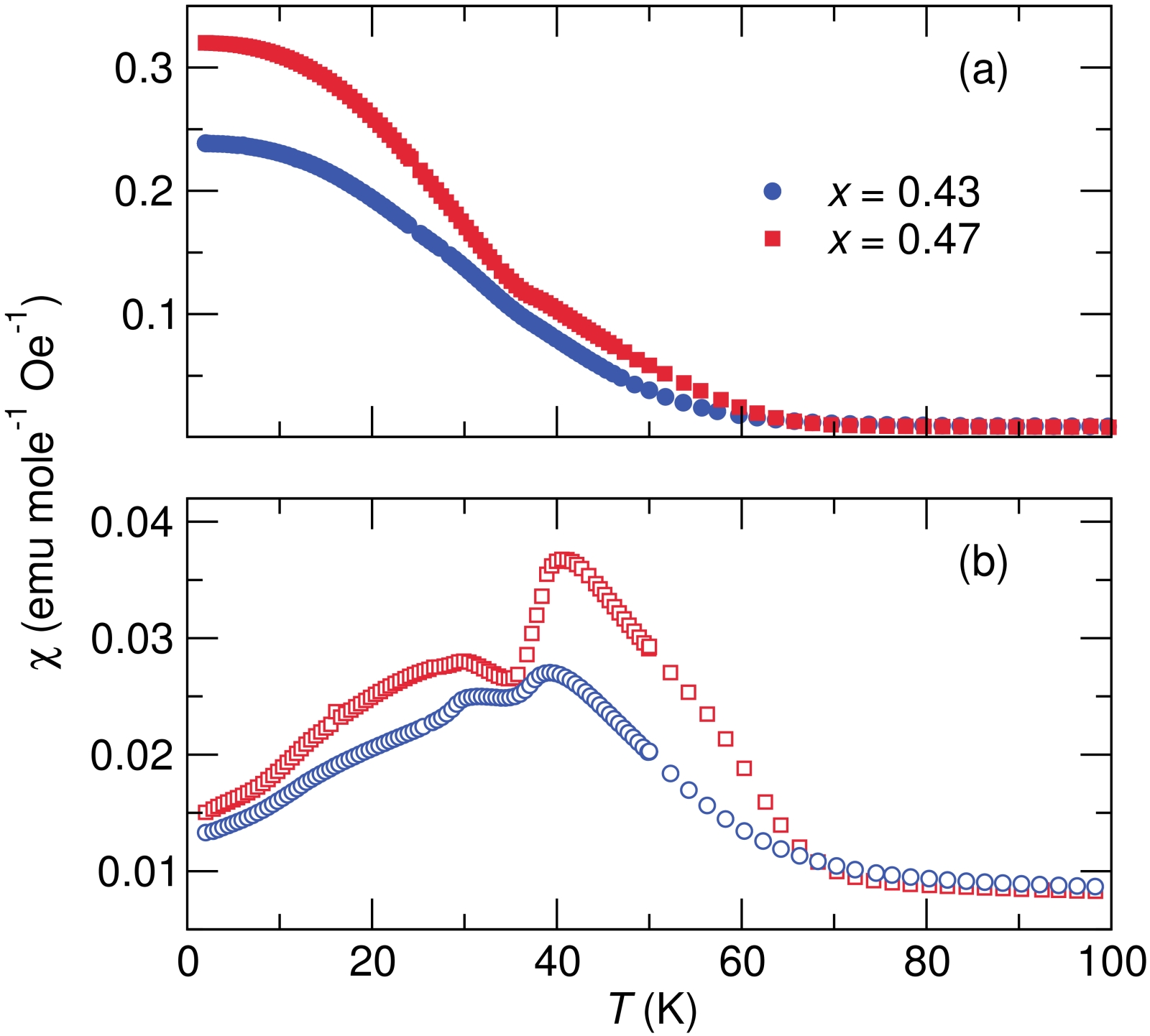} 
\caption{(a) Field-cooled and (b) zero-field-cooled magnetic susceptibility 
of the compounds with $x$ = 0.43 and $x$ = 0.47 the showing multiple 
magnetic transition temperatures present in these samples.\label{fig:cub-tet}}
\end{figure}

There are two ordering temperatures in the magnetic susceptibility of
samples $x$ = 0.43 and $x$ = 0.47 (Figure\,\ref{fig:cub-tet},
Table\,\ref{table:magnetism}). This suggests the presence of two different
kinds of long range interactions or magnetic compensation. Structural
studies of these samples showed that compositions $x$ $\leq$ 0.43
have a cubic average structure at room temperature.  Under similar
conditions, compositions $x$ $\geq$ 0.47 showed tetragonal average
symmetry. Average cubic and tetragonal symmetry was present in the range
0.43 $\leq$ $x$ $\leq$ 0.47 at 300\,K. Shoemaker \textit{et al.}
suggest that although locally CuO$_4$ tetrahedra are distorted for Cu$^{2+}$
content $\leq$ 0.43 due to the Jahn-Teller activity of tetrahedral
Cu$^{2+}$, the average structure remains cubic at 300\,K. At $x$ =
0.43, the distorted CuO$_4$ distribution is sufficient to cause a
cooperative effect and the tetragonal phase appears.\cite{Shoemaker2010}
Here, magnetic susceptibility studies complement structural studies well. We
find that in Cu$^{2+}$ concentrations $x$ $\leq$ 0.2, magnetism is
antiferromagnetic as occurs in MgCr$_2$O$_4$.  For $x$ above 0.6,
ferrimagnetism develops (Figure\,\ref{fig:chi}). Although addition of
Cu$^{2+}$ increases ferrimagnetic \textit{A}-\textit{B} interactions,
antiferromagnetism dominates at low $x$ values.  Multiple transitions
at $x$ = 0.43, suggest that \textit{A}-O-\textit{B} coupling is
sufficient to cause ferrimagnetic (Fi) long range order in addition to
antiferromagnetic (AF) order. We attribute the presence of two magnetic
transitions in the range 0.43 $\leq$ $x$ $\leq$ 0.47 to coexisting Fi
and AF interactions. Evidence of structural phase separation in the range
0.43 $\leq$ $x$ $\leq$ 0.47 at 300\,K supports the magnetic data
showing two distinct yet coexisting kinds of magnetic
order.\cite{Shoemaker2010} For $x$ $>$ 0.6, ferrimagnetism
prevails.  

\begin{figure}
\centering \includegraphics[width=0.75\textwidth]{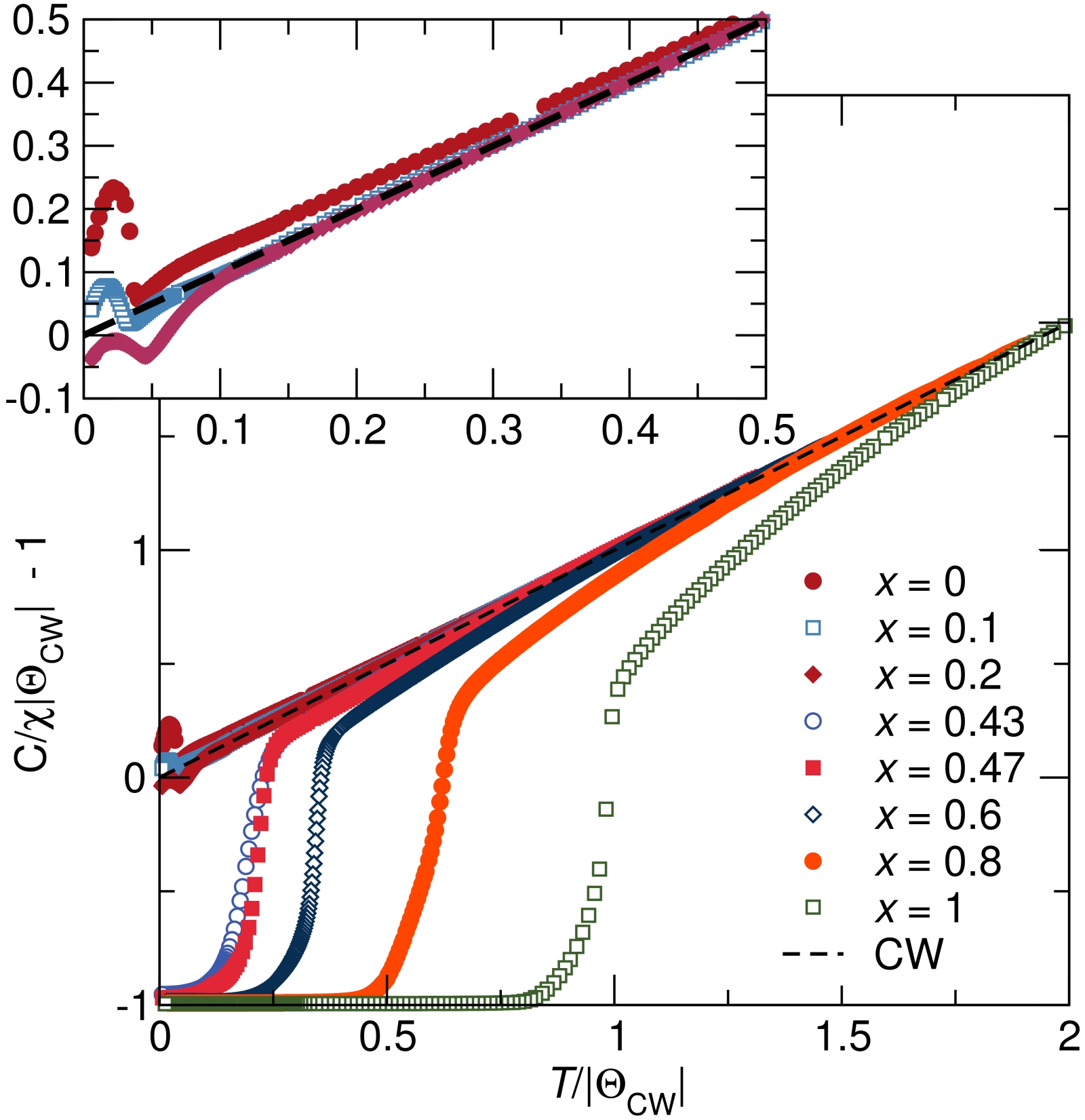} 
\caption{ \label{fig:cw} Normalized plots of inverse field-cooled magnetic 
susceptibility of Mg$_{1-x}$Cu$_x$Cr$_2$O$_4$ acquired in a 0.1\,T field, as
a function of $T/|\Theta_{CW}|$. The black dashed line represents
ideal Curie-Weiss (CW) behavior. For small $x$, the susceptibility 
follows Curie-Weiss behavior to $T \ll |\Theta_{CW}|$ and at $T_N$ deviate 
positively from ideal CW behavior as seen also in the inset. Compounds with 
$x$ = 0.43, 0.47, 0.6, 0.8, and 1 deviate negatively from ideal CW behavior 
at temperatures closer to $|\Theta_{CW}|$. \label{fig:cwfits}}
\end{figure}

High temperature susceptibility data was fit to the Curie-Weiss (CW)
equation (Equation\,\ref{cw1}) to obtain the effective paramagnetic moment
($\mu_{eff}$) and $\Theta_{CW}$. 

\begin{eqnarray}
\chi = \frac{\it{C}}{\it{T} - \Theta_{CW}}
\label{cw1}
\end{eqnarray}

The Curie constant ($C$) yields an effective paramagnetic moment
following the expression $\mu_{eff}$ = $\sqrt{3K_BC/N_A}$. $\Theta_{CW}$ is
a measure of the strength and nature of magnetic interactions. A plot of the
scaled normalized inverse susceptibility shown by Equation\,\ref{Equation:CW2} 
simplifies comparison of magnetic behavior in compounds where magnetic 
interactions evolve significantly with composition.\cite{Melot2009}

\begin{eqnarray}
\frac{C}{\chi |\Theta_{CW}|} + sgn(\Theta_{CW})= \frac{\it{T}}{|\Theta_{CW}|}
\label{Equation:CW2}
\end{eqnarray}

Inverse susceptibility scaled according to Equation {\ref{Equation:CW2}
using values obtained from Curie Weiss fits of the high temperature
susceptibility data 300 $\leq$ $T$ $\leq$ 390\,K are presented in
Figure\,\ref{fig:cwfits}. Ideal CW paramagnetism occurs in all samples when
$T_N/|\Theta_{CW}|$ $\geq$ 1 with the exception of CuCr$_2$O$_4$. In
the paramagnetic regime, spins are non-interacting. Deviations from CW
behavior denote the onset of long- or short-range interactions.
Uncompensated interactions contribute to negative deviations, while
compensated interactions lead to positive deviations. Compensated
interactions in MgCr$_2$O$_4$ gradually become uncompenstated with Cu$^{2+}$
doping. A neutron study of the magnetic structure of CuCr$_2$O$_4$ showed
that chromium sublattices yield a net moment which is partially compensated
by the Cu$^{2+}$ sublattice in the ordered state.\cite{Prince1957} It is
likely that this effect develops gradually with doping of Cu$^{2+}$ in
MgCr$_2$O$_4$. Deviations from ideal CW behaviour in CuCr$_2$O$_4$ when 
$\it{T}_N/|\Theta_{CW}|$ $\geq$ 1 indicate the presence of short range
uncompensated interactions above $\it{T}_N$ (Figure\,\ref{fig:cwfits}).  

The normalized CW plot is a direct indicator of frustrated magnetism.
$T_N/|\Theta_{CW}|$ is the inverse of the frustration parameter
($f$). The onset of long range order when $T_N/|\Theta_{CW}|$
$\ll$ 1 is a sign of strong frustration.\cite{Ramirez1994} The canonical
geometrically frustated antiferromagnet, MgCr$_2$O$_4$, shows a high
frustration index (Figure\,\ref{fig:cwfits}, Table \ref{table:magnetism}).
Suprisingly, the compound $x$ = 0.1, is the most frustrated of the
series despite the presence of random Cu$^{2+}$ ions. Possibly, dilute
Cu$^{2+}$ in Mg$_{0.9}$Cu$_{0.1}$Cr$_2$O$_4$ disrupt Cr$^{3+}$-Cr$^{3+}$
interactions inhibiting the onset of long range order and decreasing
$T_N$. Except for the composition $x$ = 0.1, Cu$^{2+}$
doping eases frustration in MgCr$_2$O$_4$ and this occurs because of the
disruption of symmetric Cr$^{3+}$-Cr$^{3+}$ interactions with random
Cu$^{2+}$ distribution in the doped compounds. Cu$^{2+}$ interferes with
\textit{B}-\textit{B} coupling by adding Cu$^{2+}$-Cr$^{3+}$ and
Cu$^{2+}$-Cu$^{2+}$ interactions. Additionally,  crystal distortions arise
in CrO$_6$ due to the Jahn-Teller effect of Cu$^{2+}$ depending on proximity
to CuO$_4$ tetrahedra. Differences in Cr-Cr distances and Cr-O-Cr angles due
to structural distortions break the degeneracy of exchange coupling routes
ultimately easing frustration.

\begin{table}
\label{table:magnetism}
\caption{Magnetic data of the series Mg$_{1-x}$Cu$_x$Cr$_2$O$_4$. 
Experimental $\mu_{eff}$ and $\Theta_{CW}$ obtained from
fitting susceptibility data in the temperature range 300 $\leq$ T $\leq$ 
390\,K to the Curie-Weiss equation. $T_N$ is taken as the temperature where
$\partial{\chi}$/$\partial{T}$ is maximum. Calculated spin-only values of
$\mu_{eff}$ are also presented.}
\begin{center}
\begin{tabular}{lllll}
\hline
\hline
$x$           & $\mu_{eff}$ ($\mu_{B}$, expt.)  & $\mu_{eff}$ ($\mu_{B}$, calc.) & $\Theta_{CW}$ (K) & $T_{N}$\\
\hline
MgCr$_2$O$_4$ & 5.42                     & 5.47                   & $-$368            & 12.5   \\
$x$ = 0.1     & 5.34                     & 5.50                   & $-$361            & 10.5   \\
$x$ = 0.2     & 5.24                     & 5.53                   & $-$329            & 16     \\
$x$ = 0.43    & 5.17                     & 5.59                   & $-$314            & 37, 27 \\
$x$ = 0.47    & 4.94                     & 5.60                   & $-$295            & 38, 29 \\
$x$ = 0.6     & 4.89                     & 5.64                   & $-$262            & 35     \\
$x$ = 0.8     & 4.54                     & 5.69                   & $-$202            & 91     \\
CuCr$_2$O$_4$ & 4.48                     & 5.74                   & $-$148            & 128    \\
\hline
\hline
\end{tabular}
\end{center}
\end{table}

\begin{figure}
\centering \includegraphics[width=0.75\textwidth]{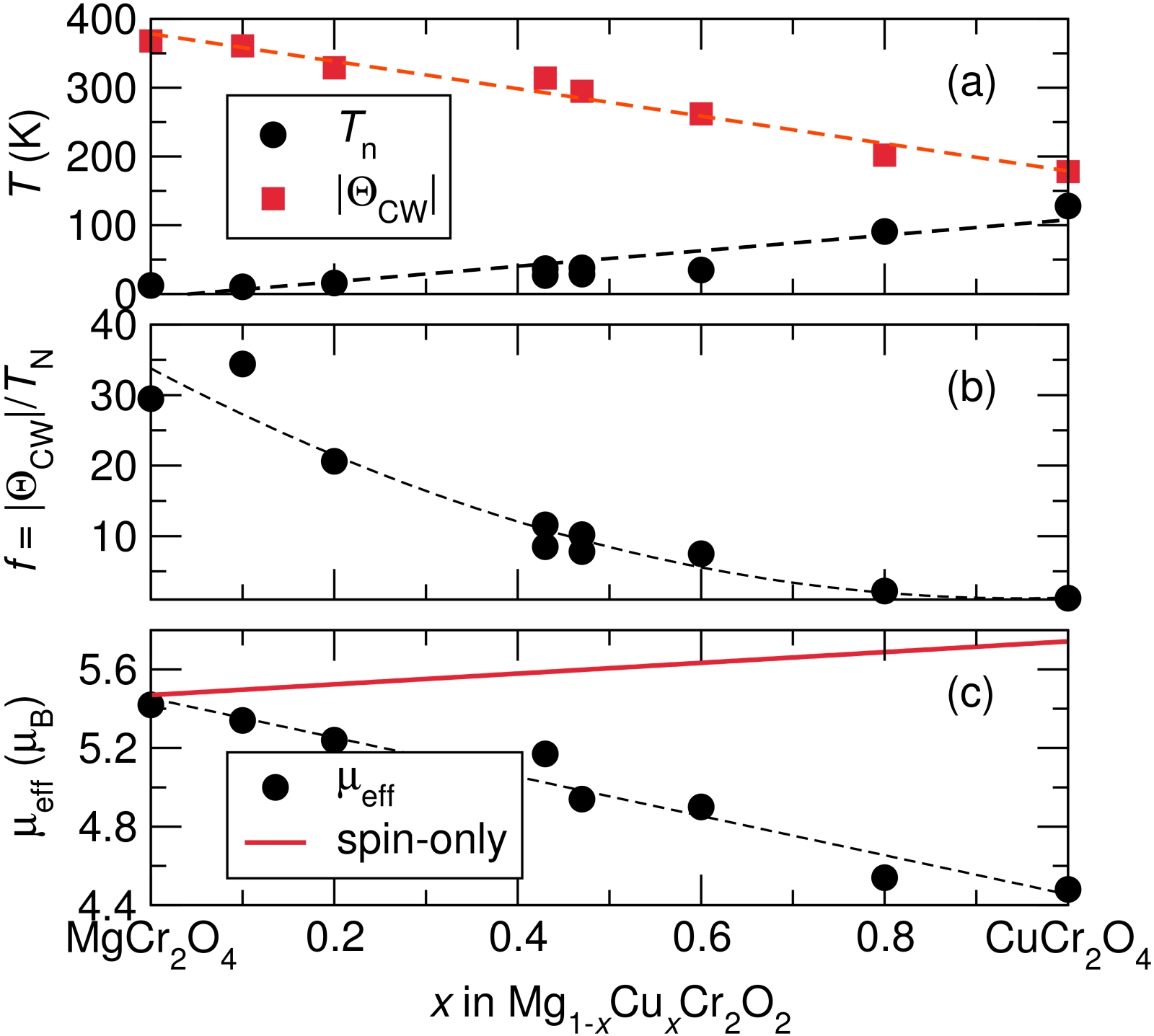} 
\caption{ \label{fig:cwdata} Trends in magnetic properties of
Mg$_{1-x}$Cu$_x$Cr$_2$O$_4$. 
(a) $T_N$ increases while $\Theta_{CW}$ decreases with $x$. (b)The 
frustration index $f = |\Theta_{CW}|/T_N$ decreases with $x$ approaching
$f$ = 1 as $x$ = 1. (c) Experimental $\mu_{eff}$ decreases with $x$ 
although the expected spin-only $\mu_{eff}$ is predicted to increase.}
\end{figure}

 Values obtained from fitting high temperature susceptibility data to the CW
equation are tabulated in Table\,\ref{table:magnetism} and plotted in 
Figure\,\ref{fig:cwdata}.  An expected spin-only effective moment of series
compositions calculated using the expression $\mu_{eff} = \sqrt{x\mu_{Cu}^2
+ 2\mu_{Cr}^2}$ is also tabulated. Here $\mu_{eff} =
2\mu_B\sqrt{\it{S}(\it{S}+1)}$ where S = 1/2 and 3/2 for Cu$^{2+}$ and
Cr$^{3+}$ respectively gives the expected effective moment from Cu$^{2+}$
and Cr$^{3+}$. The effective moment is expected to increase with Cu$^{2+}$
concentration. Suprisingly, a decrease in $\mu_{eff}$ occurs with Cu$^{2+}$
doping. We speculate that short range interactions develop in the
paramagnetic regime in Cu$^{2+}$ rich samples leading to the
underestimation of the spin-only effective moment. The orbital contribution
to the effective magnetic moment in octahedral Cr$^{3+}$ and Jahn-Teller
active Cu$^{2+}$ is expected to be heavily quenched, hence, the orbital
moment is not considered.

A negative theoretical ordering temperature ($\Theta_{CW}$) is determined
for all samples, confirming that spin coupling is antiferromagnetic or
ferrimagnetic. The magnitude of $\Theta_{CW}$ decreases with Cu$^{2+}$
content, which is unexpected as Cu$^{2+}$ increases the number of
interactions. A similar trend in Curie-Weiss temperature has been reported
in the systems Zn$_{1-x}$Cu$_x$Cr$_2$O$_4$ \cite{Wang2008} and
Cd$_{1-x}$Cu$_x$Cr$_2$O$_4$.\cite{Wang2007} The earlier works postulate
that the decrease is due to ferromagnetic interactions from the Cu$^{2+}$
sublattice. Geometry and competition between magnetic interactions results
in weaker overall interactions in Cu$^{2+}$ rich samples. 

Despite the increase in $\it{T}_N$ with Cu$^{2+}$, the theoretical ordering
temperature ($\Theta_{CW}$) shows an opposite trend. $\Theta_{CW}$
decreases with Cu$^{2+}$ concentration. The combined change in
$\Theta_{CW}$ and $\it{T}_N$ results in a dramatic decrease of the
frustration parameter with Cu$^{2+}$ content. In agreement with previous
similar studies\cite{Melot2009, Wang2008, Wang2007}, we find that
frustration is strongly quenched with the addition of spins in the
\textit{A} sublattice. These results show that lattice geometry does not
prevent spin order in Cu$^{2+}$ doped samples. The sample $x$ = 0.1
exhibits a curious trend. In this compound, the  frustration index
increases despite Cu$^{2+}$ doping. This was also observed in the study of
magnetism of the series Zn$_{1-x}$Co$_x$Cr$_2$O$_4$.\cite{Melot2009} It is
likely that dilute randomly distributed Cu$^{2+}$ introduces disorder in
the exchange coupling routes thus disrupting long range order. The system
assumes a frozen disordered state at the suppressed transition
temperature.

\begin{figure}
\centering \includegraphics[width=0.75\textwidth]{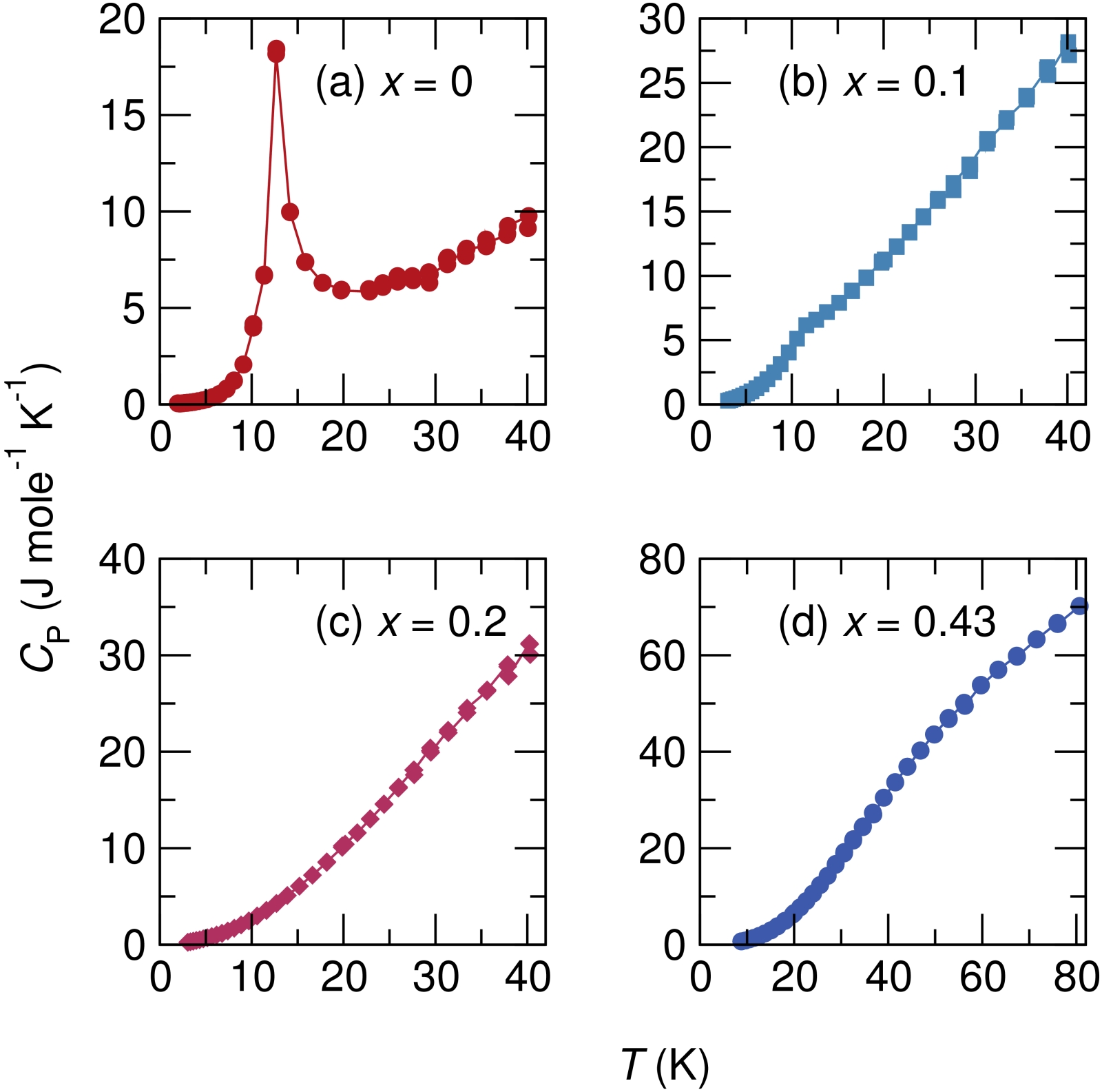} 
\caption{ \label{fig:HC} Temperature dependent heat capacity of compounds with
$x$ = 0, 0.1, 0.2, and 0.43 in Mg$_{1-x}$Cu$_x$Cr$_2$O$_4$ under zero
field. (a) Heat capacity of the sample $x$ = 0 showing a sharp peak at 12.5\,K;
(b) A broader peak occurs at 11.5\,K in the sample $x$ = 0.1; 
(c) In the samples with $x$ = 0.2, a smooth decrease in heat capacity with 
temperature is observed (d) The sample with $x$ = 0.43 displays a broad 
bulge in the heat capacity at $\approx$ 50\,K.}
\end{figure}

Entropy changes are well described by specific heat measurements. A sharp
peak in the specific heat of MgCr$_2$O$_4$ indicates changes in entropy at
$T$ $\approx$ 12.5\,K [Figure\,\ref{fig:HC}(a)]. The transition
temperature recorded in heat capacity coincides with the magnetic ordering
temperature signalling the onset of long range AF order. A broad shallow
peak at $T$ $\approx$ 11\,K appears at the magnetic ordering
temperature of the sample $x$ = 0.1 suggesting that there is a
remanent magnetic entropy below the ordering temperature 
[Figure\,\ref{fig:HC}(b)]. 
The suggestion that for dilute Cu$^{2+}$ concentrations,
the spins freeze in a disordered state rather than one with long range
order, is further supported by the absence of an anomaly in the heat
capacity of the sample $x$ = 0.2 [Figure\,\ref{fig:HC}(c)].  In the
compound $x$ = 0.43, a broad hump develops at 30\,K $\leq$ $T$
$\leq$ 70\,K. The broad range in temperature of the transition indicates a
more diffuse magnetic transition and the onset of long range order at
$x$ $\geq$ 0.43 [Figure\,\ref{fig:HC}(d)].   

\begin{figure}
\centering \includegraphics[width=0.75\textwidth]{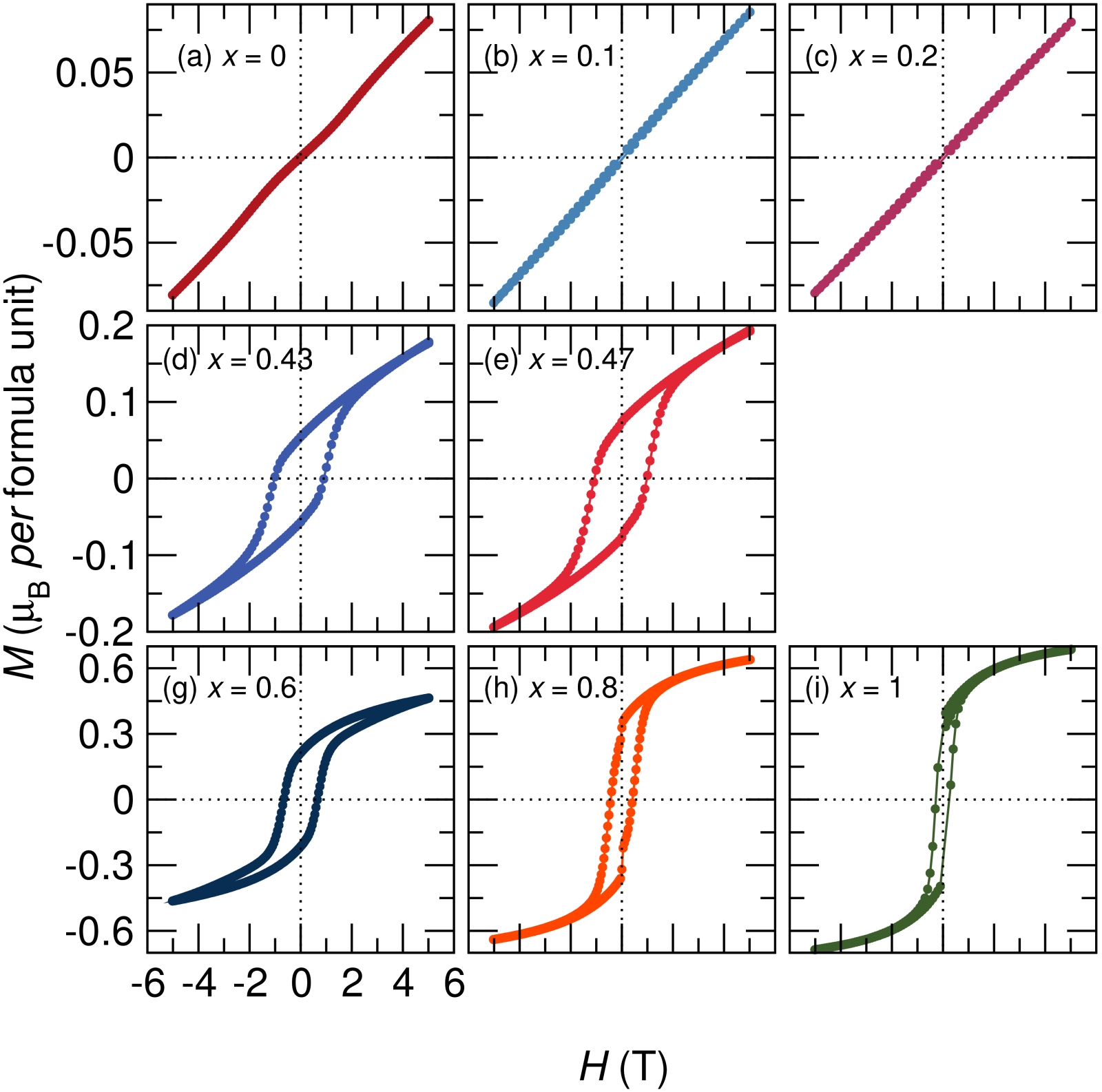} 
\caption{ \label{fig:MH} Isothermal magnetization traces of the different
Mg$_{1-x}$Cu$_x$Cr$_2$O$_4$ compounds, measured at 2\,K except for 
the samples with $x$ = 0.6 and 0.8 which were measured at 5\,K.
The compound (a) $x$ = 0 shows a linear variation of magnetization with field. 
Magnetic anisotropy induces slight nonlinearity at $\pm$ 1.5\,T as reported by 
Suzuki \textit{et al.}\cite{Suzuki2007} In (b) $x$ = 0.1 and (c) $x$ = 0.2
linear dependence of the magnetization on field is again observed. In (d), the
$x$ = 0.43 sample and in (e) the $x$ = 0.47 sample display a larger increase in
magnetization as compared to samples with $x$ = 0, 0.1, and 0.2. A
large coercive field develops as observed from the wide hysteresis
loops. At high fields the magnetization does not saturate. The samples
with (f) $x$ = 0.6 and (g) 0.8 are ferrimagnetic with open hysteresis loops
and almost saturating magnetization. (h) Field dependence of the 
magnetization of CuCr$_2$O$_4$ at 2\,K showing well behaved hysteresis
and an almost saturated magnetization of 0.7 $\mu_B$ per formula unit with
small coercive field.}
\end{figure}

A progressive transition from antiferromagnetic to ferrimagnetic order is
observed in isothermal field dependent magnetization measurements (Figure
\ref{fig:MH}). In MgCr$_2$O$_4$, a linear dependence of the magnetization
on field that is characteristic of antiferromagnetic order is observed.
Slight nonlinearity in the field dependent magnetization develops at
$\pm$1.5 T in MgCr$_2$O$_4$ [Figure \ref{fig:MH}(a)]. This nonlinearity
originates from a field-driven change in local magnetic structure as
postulated by Suzuki \textit{et. al}.\cite{Suzuki2007}. Antiferromagnetism
persists in samples $x$ = 0.1 and 0.2 [Figure\,\ref{fig:MH}(b,c)],
however, coercivity develops as the loops are open upon close examination.
Coercivity is highest in samples $x$ = 0.43 and 0.47 
[Figure\,\ref{fig:MH}(d,e)]. Additionally, earlier studies on exchange bias 
effects in these compounds revealed an increase in the exchange bias field
($H_e$) with Cu$^{2+}$ in samples $x$ = 0.1, 0.2 and
0.43.\cite{Shoemaker2010} Antiferromagnetic-ferrimagnetic interfaces are a
common cause of enhanced coercivity and shifted
hysteresis.\cite{Schuller2000,Schuller1999} Given that the samples
$x$ = 0.43 and $x$ = 0.47 bridge the AF and Fi regions, we propose that
pinning of spins at the Fi-AF cluster interfaces contributes to the
enhanced coercivity. 

Coupled with the increased coercivity in compounds $x$ = 0.43 and
$x$ = 0.47 is the change from AF to Fi behavior that is evident in field- and
temperature-dependent behavior. In samples with $x$ = 0.6, $x$ = 0.8, and
$x$ = 1, the coercivity decreases and the field dependent magnetization becomes
ferrimagnetic. CuCr$_2$O$_4$ has a coercive field(\textit{H$_e$\textit{}})
of about 0.25 T and a saturation moment of 0.68 $\mu_B$ per formula
unit. The N\'{e}el model of  antiferromagnetism predicts a saturation
moment of 5 $\mu_B$ per formula unit in CuCr$_2$O$_4$. Since the predicted
moment far exceeds the measured value, neutron diffraction measurements in
the ordered state were used to resolve this discrepancy.\cite{Prince1957} A
triangular arrangement of spins explains the low moment. Cr$^{3+}$ on the
001 planes are alligned parallel and opposite to Cr$^{3+}$ in adjacent
planes yielding a net moment from the Cr$^{3+}$ sublattices. The Cu$^{2+}$
sublattice couples antiferromagnetically to the net moment of the Cr$^{3+}$
sublattices creating a triangular configuration of spins. A composition
dependent saturation magnetization is observed in samples $x$ =
0.43, 0.47, 0.6, 0.8, and 1. The saturation moment increases with
Cu$^{2+}$. 

We can assemble the (Mg,Cu)Cr$_2$O$_4$ magnetic phase diagram in Figure
\ref{fig:phase}  by combining magnetic ordering transitions and heat
capacity measurements for the various compositions in this study.
Transition temperatures specified by anomalies in susceptibity and heat
capacity measurements are used to demarcate phase boundaries.  
MgCr$_{2}$O$_4$ enters a long range AF state at $T$ = 12.5\,K marked
by sharp anomalies in both susceptibility and heat capacity. The sharp cusp
in the susceptibility of MgCr$_{2}$O$_4$ is replaced with a rounded peak in
compositions $x$ = 0.1 and 0.2. Although the magnetic order in these
compositions is antiferromagnetic as evidenced by the linear
dependence of magnetization on field, they possess significant disorder.
Disorder is indicated by the suppressed anomaly in heat capacity of the
sample $x$ = 0.1 and the complete absence of a transition in the
specific heat of $x$ = 0.2. All compositions 0.2 $\leq$ x $\leq$ 0.6
exhibit glassy magnetic states. In the sample $x$ = 0.6,
ferrimagnetism develops indicated by the similarities in field and
temperature dependent magnetization between the compound and the end-member
CuCr$_2$O$_4$. All compositions are paramagnetic above the magnetic
transition temperatures. 

\begin{figure}
\centering \includegraphics[width=0.75\textwidth]{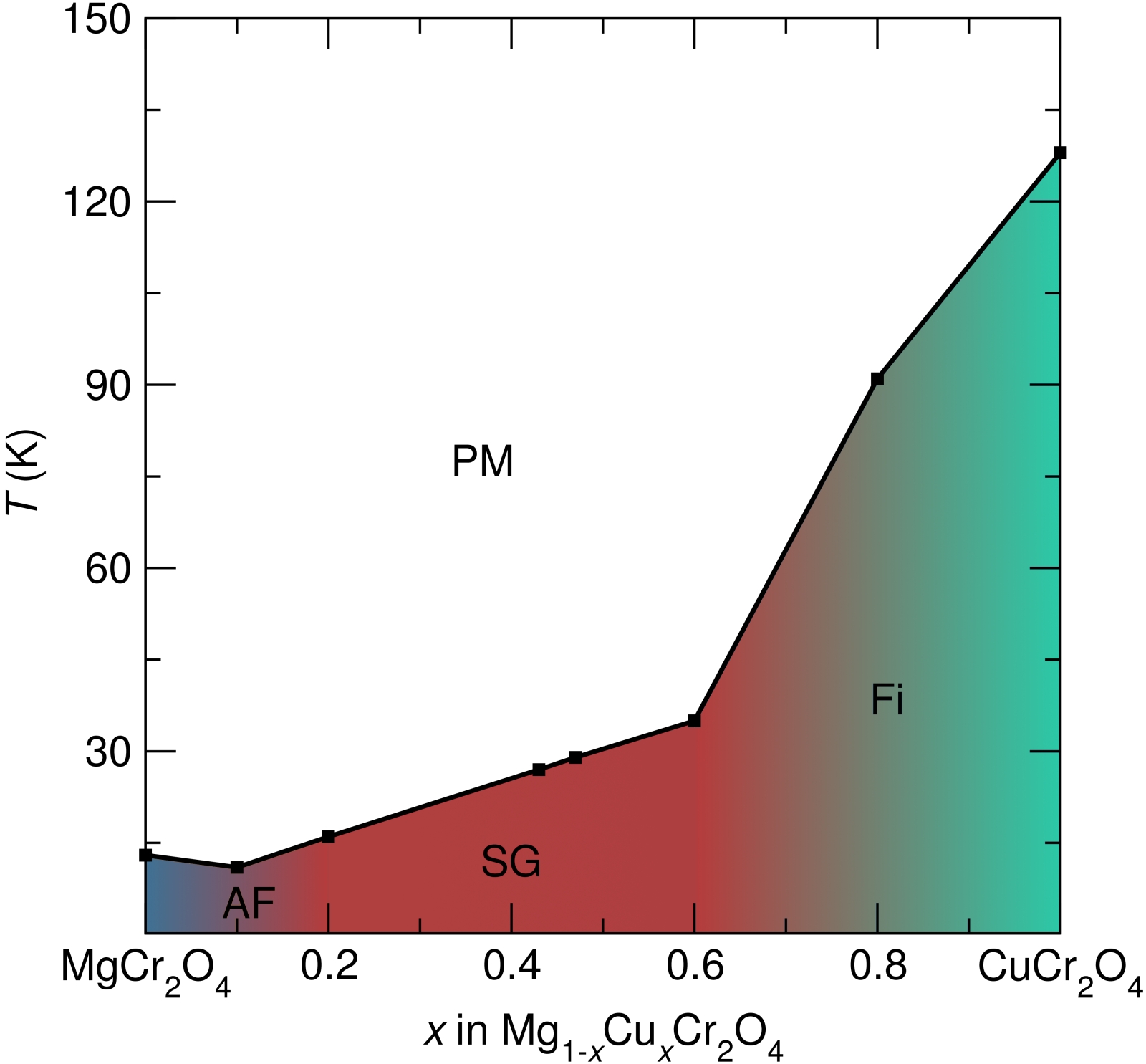} 
\caption{Suggested magnetic phase diagram for the solid solution
Mg$_{1-x}$Cu$_x$Cr$_2$O$_4$ based on susceptibility, heat capacity 
and magnetization measurements. In compounds with $x$ $\leq$ 0.2, the magnetic 
ordering is antiferromagnetic (AF) below $T_N$. For compositions 
with 0.2 $\leq$ $x$ $\leq$ 0.6, glassy behaviour (SG) is observed, and 
for $x$ larger than 0.6, the ordering is largely ferrimagnetic (Fi), 
albeit not N\'eel-like.
\label{fig:phase}}
\end{figure}

\section{Conclusions}

Recent years have seen renewed focus on frustrated magnetic systems as
a consequence of the recognition that they can give rise to exotic 
magnetic ground states.\cite{Ramirez1994,Ueda2006,Chalker1998} 
The introduction of spins in
the \textit{A} site of geometrically frustrated spinels has been suggested
as a means of relieving frustration.\cite{Melot2009, Wang2008, Wang2007}.
We find in agreement with this proposition, that Cu$^{2+}$ alieviates
frustration in all Mg$_{1-x}$Cu$_x$Cr$_2$O$_4$ compositions studied with
the exception of $x$ = 0.1.
Temperature- and field-dependent magnetization indicate a largely 
antiferromagnetic ground state in the compositions $x$ $<$ 0.2, 
disordered glassy states with open hysteresis loops in the range 
0.2 $\leq x \leq$ 0.6, and largely ferrimagnetic ground states for $x$ $>$ 0.6, 
albeit with significantly smaller saturation magnetizations than would be
found for N\'eel -ordered states. Specific heat measurements support these
conclusions for samples with small $x$.
Emergent behavior at the AF/Fi phase boundary, indicated by large
coercive fields and multiple magnetic transition temperatures, suggests
microscopic interactions between AF and Fi clusters.

\section{Acknowledgements} 
MCK thanks Joshua Kurzman and Phillip Barton for helpful comments. We gratefully
acknowledge support from the National Science Foundation through a Materials 
World Network grant (DMR 0909180). MCK is supported by the Schlumberger 
Foundation Faculty for the Future Fellowship. SLM is supported by the RISE 
program at the Materials Research Laboratory. Use of shared experimental 
facilities of the Materials Research Laboratory: an NSF MRSEC, supported by 
NSF DMR 1121053 is acknowledged. The MRL is a member of the the NSF-supported 
Materials Research Facilities Network (www.mrfn.org). The 11-BM beamline at 
the Advanced Photon Source is supported by the Department of Energy, Office of 
Science, Office of Basic Energy Sciences, under Contract No. DE-AC02-06CH11357.

\clearpage


\medskip

\begin{thebibliography}{10}

\bibitem{Thackeray1987}
Thackeray M M, de Picciotto L A, de Kock A, Johnson P J, Nicholas V A, and Adendorff K T 1986 {\em J. Power Sources\/ } {\bf 21} 1

\bibitem{Yamasaki2006}
Yamasaki Y, Miyasaka S, Kaneko Y, He J-P, Arima T, and Tokura Y 2006 {\em Phys. Rev. Lett.\/ } {\bf 96} 207204 

\bibitem{Lawes2006}
Lawes G, Melot B, Page K, Ederer C, Hayward M A, Proffen Th, and Seshadri R 2006 {\em Phys. Rev. B\/ } {\bf 74} 024413 

\bibitem{Gao2009}
Zhu J, and Gao Q 2009 {\em Microporous Mesoporous Mater.\/ } {\bf 124} 144  

\bibitem{Yaresko2008}
Yaresko A N 2008 {\em Phys. Rev. B.\/ } {\bf 77} 115106 

\bibitem{Loidl2009}
Glazkov V N, Farutin A M, Tsurkan V, Krug von Nidda H-A and Loidl A 2009 {\em Phys. Rev. B\/ } {\bf 79} 021131 

\bibitem{Takagi2005} 
Ueda H, Katori H A, Mitamura H, Goto T, and Takagi H 2005 {\em Phys. Rev. Lett.\/ } {\bf 94} 047202 

\bibitem{Cheong2002}
Lee S-H, Broholm C, Ratcliff W, Gasparovic G, Huang Q, Kim T H, and Cheong S-W 
2002 {\em Nature\/ } {\bf 418} 856 

\bibitem{Balents2010}
Balents L 2010 {\em Nature\/ } {\bf 466} 216007 

\bibitem{Ramirez2006}
Moessner R and Ramirez A P 2006 {\em Phys. Today\/ } {\bf 59} 24

\bibitem{Orgel1957}
Dunitz J D and Orgel L E 1957 {\em J. Phys. Chem. Solids\/ } {\bf 3} 318

\bibitem{Miller1959}
Miller A 1959 {\em J. Appl. Phys.\/ } {\bf 30} 4 

\bibitem{Anderson1956}
Anderson P W 1956 {\em Phys. Rev. \/ } {\bf 102} 4 

\bibitem{Cheong2007}
Lee S- H, Gasparovic G, Broholm C, Matsuda M, Chung J-H, Kim Y J, Ueda H, Xu G, Zschack P, Kakurai K, Takagi H, Ratcliff W, Kim T H, and Cheong S-W 2007 {\em J. Phys.: Condens. Matter\/ } {\bf 19} 145259 

\bibitem{Chalker1997}
Moessner R and Chalker J T 1998 {\em Phys. Rev. Lett.\/ } {\bf 80} 13 

\bibitem{Sondhi2002}
Tchernyshyov O, Moessner R, and Sondhi S L 2002 {\em Phys. Rev. Lett.\/ } {\bf 88} 067203 

\bibitem {Loidl2010}
Kant C, Deisenhofer J, Tsurkan V, and Loidl A 2010 {\em J. Phys: Conf. Ser.\/} {\bf 200} 032032 

\bibitem{Seshadri2006} 
Lawes G, Melot B, Page K, Ederer C, Hayward M A, Proffen Th, and Seshadri R 2006 {\em Phys. Rev. B\/} {\bf 74} 024413

\bibitem{Ederer2007} 
Ederer C and Komelj M 2007 {\em Phys. Rev. B\/} {\bf 76} 064409

\bibitem{Prince1957}
Prince E 1957 {\em Acta Crystallogr. \/ } {\bf 10} 554 

\bibitem {Pickart1964}
Shirane G, Cox D E, and Pickart S J 1964 {\em J. Appl. Phys.\/} {\bf 35} 3 

\bibitem {Suzuki2004}
Tomiyasu K, Fukunaga J, and Suzuki H 2004 {\em Phys. Rev. B\/} {\bf 70} 214434 

\bibitem{Tokura2009} 
Bord\'{a}cs S, Varjas D, K\'{e}zsmarki I, Mih\'{a}ly G, Balassarre L, Abouelsayed A, Kuntscher C A, Ohgushi K, and Tokura Y 2009 {\em Phys. Rev. Lett.\/} {\bf 103} 077205

\bibitem{Tsunoda2007}
Suzuki H and Tsunoda Y 2007 {\em J. Phys. Chem. Solids\/ } {\bf 68} 2060 

\bibitem{Cava2011}
Dutton S E, Huang Q, Tchernyshyov O, Broholm C L, and Cava R J 
2011 {\em Phys. Rev. B\/ } {\bf 83} 064407 

\bibitem{Attfield2008}
Ortega-San-Martin L, Williams A J, Gordon C D, Klemme S, and Attfield J P 
2008 {\em J. Phys.: Condens. Matter\/ } {\bf 20} 104238 

\bibitem{Baehtz2002}
Ehrenberg H, Knapp M, Baehtz C and Klemme S 2002 {\em Powder Diffraction\/ } {\bf 17} 230 

\bibitem{Gmelin2000}
Klemme S, O'Neill H, Schnelle W, and Gmelin E 2000 {\em Am. Mineral.\/ } {\bf 85} 1686 

\bibitem{Kanamori1960}
Kanamori J 1960 {\em J. Appl. Phys.\/ } {\bf 31} S14 

\bibitem{Murthy1983}
Murthy K S R C, Ghose J, and Rao E N 1983 {\em J. Mater. Sci. Lett.\/ } {\bf 2} 393 

\bibitem{Gerloch1981}
Gerloch M 1981 {\em Inorg. Chem.\/ } {\bf 20} 638 

\bibitem{O'Neill1997}
Dollase W A, and O'Neill H St. C 1997 {\em Acta Crystallogr., Sect. C: Cryst. Struct.
Commun. } {\bf C53} 657 

\bibitem{Shoemaker2010}
Shoemaker D P and Seshadri R 2010 {\em Phys. Rev. B\/} {\bf 82} 214107 

\bibitem{Melot2009}
Melot B, Drewes J E, Seshadri R, Stoudenmire E M, and Ramirez A P 2009 {\em J. Phys.: Condens. Matter\/ } {\bf 21} 216007 

\bibitem{Shoemaker2009}
Shoemaker D P, Rodriguez E E, Seshadri R, Abumohor I V, and Proffen Th 2009 {\em Phys. Rev. B\/} {\bf 80} 144422 

\bibitem{Broholm2010}
Lee S-H, Takagi H, Louca D, Matsuda M, Ji S, Ueda H, Ueda Y, Katsufuji T, Chung J-H, Park S, Cheong S-W, and Broholm C 2010 {\em J. Phys. Soc. Jpn.\/ } {\bf 79} 011004 

\bibitem{Ramirez1994}
Ramirez A P 1994 {\em Annu. Rev. Mater. Sci.\/} {\bf 24} 1994 453 

\bibitem{Wang2008}
Yan L-Q, Macia F, Jiang Z-W, Shen J, He L-H, and Wang F-W. 2008 {\em J. Phys.: Condens. Matter\/} {\bf 20} 255203 

\bibitem{Wang2007}
Yan L-Q, Jiang Z-W, Peng X D, He L-H, and Wang F-W 2007 {\em Powder Diffraction \/} {\bf 22} 340 

\bibitem{Schuller2000}
Leighton C, Nogu\'es J, J\~{o}nsson-\.{A}kerman B J, and Schuller I K 2000 {\em Phys. Rev. Lett.\/} {\bf 84} 3466 

\bibitem{Schuller1999}
Nogu\'es J and Schuller K I 1999 {\em J. Magn. Magn. Mater.\/} {\bf 192} 203 

\bibitem{Ueda2006}
Zhang Z, Louca D, Visinoiu A, Lee S-H, Thompson J D, Proffen Th, Llobet A, Qiu Y, Park S, Ueda Y 2006 {\em Phys. Rev. B\/} {\bf 74} 014108 

\bibitem{Chalker1998}
Moessner R and Chalker J T 1998 {\em Phys. Rev. B\/} {\bf 58} 12042

\end{thebibliography}
\end{document}